\documentclass[aps,twocolumn,superscriptaddress,prl,floatfix,british]{revtex4-1}

\usepackage{amssymb}
\usepackage{amsmath}
\usepackage{psfrag}
\usepackage{topcapt}
\usepackage{hyperref}
\usepackage[sort&compress]{natbib}
\usepackage[utf8]{inputenc}

\newcommand{\bra}[1]{|#1\rangle}

\newcommand{\ghz}{~\mathrm{GHz}}

\newcommand{\ns}{~\mathrm{ns}}

\newcommand{\wcav}{\omega_{\mathrm{c}}}

\newcommand{\EJqmax}{E_{\mathrm{J}q}^{\mathrm{max}}}

\newcommand{\ECq}{E_{\mathrm{C}q}}
\newcommand{\EJq}{E_{\mathrm{J}q}}

\newcommand{\MHz}{\mathrm{MHz}}
\newcommand{\GHz}{\mathrm{GHz}}
\newcommand{\us}{\mu\mathrm{s}}

\usepackage{ifpdf}

\ifpdf
\usepackage[pdftex]{graphicx}
\else
\usepackage{graphicx}
\fi

\begin{document}
\title{Realization of Three-Qubit Quantum Error Correction with Superconducting Circuits}

\author{M. D. Reed}
\affiliation{Departments of Physics and Applied Physics, Yale University, New Haven, Connecticut 06520, USA}
\author{L. DiCarlo}
\affiliation{Kavli Institute of Nanoscience, Delft University of Technology, Delft, The Netherlands}
\author{S. E. Nigg}
\affiliation{Departments of Physics and Applied Physics, Yale University, New Haven, Connecticut 06520, USA}
\author{L. Sun}
\affiliation{Departments of Physics and Applied Physics, Yale University, New Haven, Connecticut 06520, USA}
\author{L. Frunzio}
\affiliation{Departments of Physics and Applied Physics, Yale University, New Haven, Connecticut 06520, USA}
\author{S. M. Girvin}
\affiliation{Departments of Physics and Applied Physics, Yale University, New Haven, Connecticut 06520, USA}
\author{R. J. Schoelkopf}
\affiliation{Departments of Physics and Applied Physics, Yale University, New Haven, Connecticut 06520, USA}

\date{\today}

\ifpdf
\DeclareGraphicsExtensions{.pdf, .jpg, .tif}
\else
\DeclareGraphicsExtensions{.eps, .jpg}
\fi

\maketitle

{\bf
Quantum computers promise to solve certain problems exponentially faster than possible classically but are challenging to build because of their increased susceptibility to errors.  Remarkably, however, it is possible to detect and correct errors without destroying coherence by using quantum error correcting codes \cite{Shor1995b}.  The simplest of these are the three-qubit codes, which map a one-qubit state to an entangled three-qubit state and can correct any single phase-flip or bit-flip error of one of the three qubits, depending on the code used \cite{Nielsen2000}.  Here we demonstrate both codes in a superconducting circuit by encoding a quantum state as previously shown \cite{DiCarlo2010, Neeley2010}, inducing errors on all three qubits with some probability, and decoding the error syndrome by reversing the encoding process.  This syndrome is then used as the input to a three-qubit gate which corrects the primary qubit if it was flipped.  As the code can recover from a single error on any qubit, the fidelity of this process should decrease only quadratically with error probability.  We implement the correcting three-qubit gate, known as a conditional-conditional NOT (CCNot) or Toffoli gate, using an interaction with the third excited state of a single qubit, in $\mathbf{63~ns}$.  We find $\mathbf{85} \boldsymbol{\pm} \boldsymbol{1}\boldsymbol{\%}$ fidelity to the expected classical action of this gate and $\mathbf{78} \boldsymbol{\pm} \boldsymbol{1}\boldsymbol{\%}$ fidelity to the ideal quantum process matrix.  Using it, we perform a single pass of both quantum bit- and phase-flip error correction with $\mathbf{76} \boldsymbol{\pm} \boldsymbol{0.5}\boldsymbol{\%}$ process fidelity and demonstrate the predicted first-order insensitivity to errors.  Concatenating these two codes and performing them on a nine-qubit device would correct arbitrary single-qubit errors.  When combined with recent advances in superconducting qubit coherence times \cite{Paik2011, Kim2011}, this may lead to scalable quantum technology.
}

Quantum error correction relies on detecting the presence of errors without gaining knowledge of the encoded quantum state.  In the three-qubit code, the subspace of the two additional ``ancilla'' qubits uniquely encodes which of the the four possible single-qubit errors has occurred, including the possibility of no flip.  Critically, errors consisting of finite rotations can also be corrected by projecting this syndrome, essentially forcing the system to ``decide'' if a full phase- or bit-flip occurred \cite{Nielsen2000}.  Previous works implementing error correcting codes in liquid- \cite{Cory1998, Knill2001, Boulant2005} and solid-state \cite{Moussa2011} NMR and with trapped ions \cite{Chiaverini2004, Schindler2011} have demonstrated two possible strategies for using the error syndromes.  The first is to measure the ancillas and use a classical logic operation to correct the detected error.  This ``feed-forward'' capability is challenging in superconducting circuits as it requires a fast and high-fidelity quantum non-demolition measurement, but is likely a necessary component to achieve scalable fault-tolerance \cite{Shor1996, Nielsen2000}.  The second strategy, as recently demonstrated with trapped ions \cite{Schindler2011} and used here, is to replace the classical logic with a quantum CCNot gate which performs the correction coherently, leaving the entropy associated with the error in the ancilla qubits.  The CCNot performs exactly the action that would follow the measurement in the first scheme: flipping the primary qubit if and only if the ancillas encode the associated error syndrome.

The CCNot gate is also vital for a wide variety of applications such as Shor's factoring algorithm \cite{Shor1995} and has attracted significant experimental interest with recent implementations in linear optics \cite{Lanyon2008}, trapped ions \cite{Monz2009}, and superconducting circuits \cite{Fedorov2011, Mariantoni2011}.  Here we use the circuit quantum electrodynamics architecture \cite{Wallraff2004} to couple four transmon qubits \cite{Schreier2008} to a single microwave cavity bus \cite{Majer2007}, where each qubit transition frequency can be controlled on nanosecond timescales with individual flux bias lines \cite{DiCarlo2009} and collectively measured by interrogating transmission through the cavity \cite{Reed2010b}.  (The details of the device can be found in the Methods Summary and in Ref.~\citenum{DiCarlo2010}.)  Qubits are tuned to $6$, $7$, and $7.85 \ghz$, with the fourth at $\sim13\ghz$ and unused (hereafter referred to as $Q_1$-$Q_4$).  In this paper, we first demonstrate the three-qubit interaction used in the gate, which is the logical extension of interactions used in previous two-qubit gates \cite{DiCarlo2009, DiCarlo2010, Strauch2003}, and demonstrate how this interaction can be used to create the desired CCNot.  We then characterize its classical and quantum action and finally use the gate to demonstrate three-qubit error correction.

\begin{figure}
	\centering
	\includegraphics[scale=1]{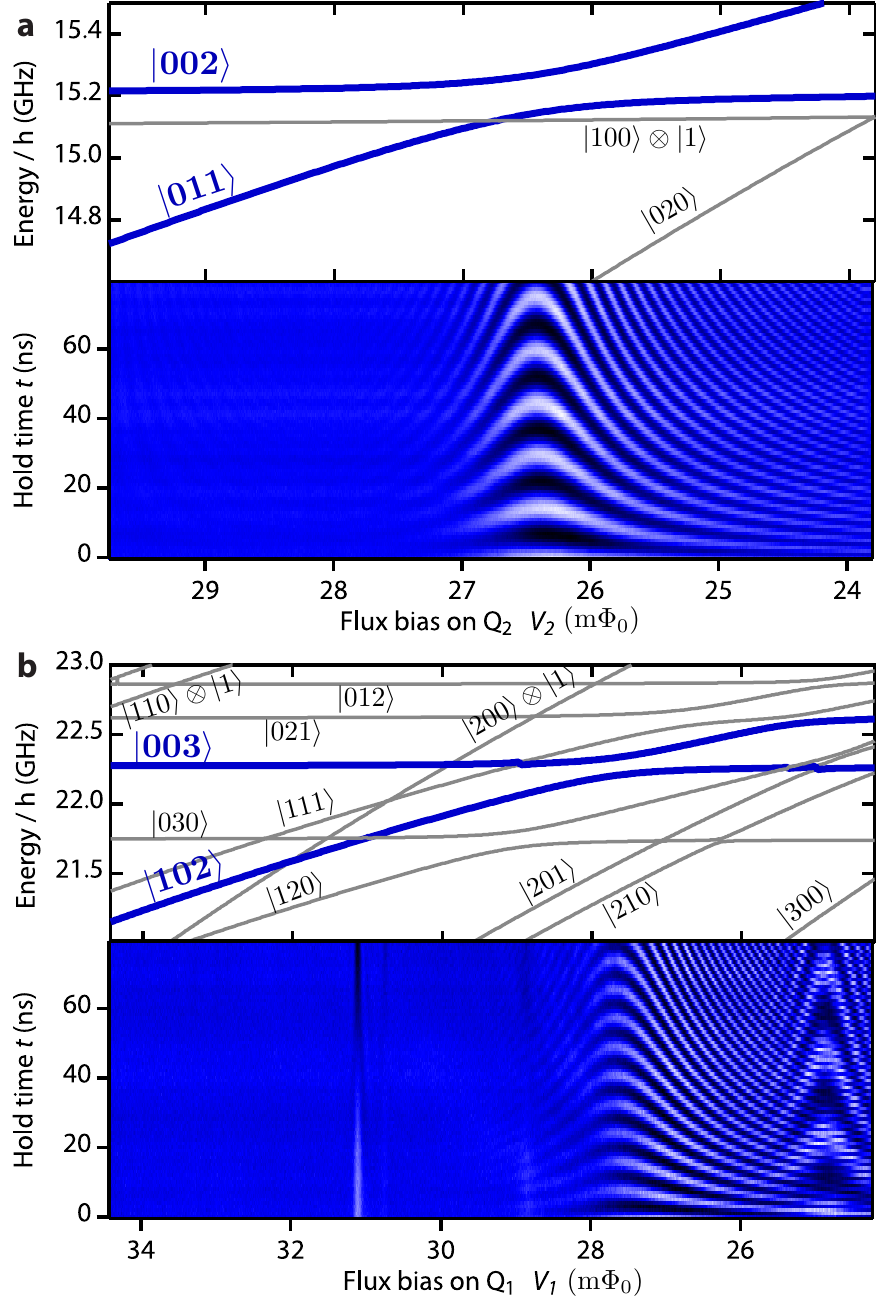}
	\caption{
		\textbf{Calculated energy spectra and time domain measurements of the interactions used in the three-qubit gate.}  
		(a) The energy spectrum of doubly excited states showing the avoided crossing between $|011\rangle$ and $|002\rangle$ (identical to that between $|111\rangle$ and $|102\rangle$ except for a $6\ghz$ offset) is shown with both (top) a numerical diagonalization of the system Hamiltonian and (bottom) a time-domain measurement as a function of the flux bias on $Q_2$.  (top) The frequencies for the involved eigenstates are blue and non-interacting eigenstates of similar energy are grey.  The notation $|abc\rangle \otimes |d\rangle$ indicates the excitation level of each qubit and the cavity photon number, respectively.  When omitted, $d=0$.  (bottom)  The state $|011\rangle$ is prepared and a square flux pulse of duration $t$ and amplitude $V_2$ is applied.  Coherent oscillations produce a ``chevron'' pattern, with darker colors corresponding to population left in $\bra{002}$.
		(b) The spectrum of triply excited states showing the avoided crossing between $|102\rangle$ and $|003\rangle$ as a function of the flux bias on $Q_1$ is characterized in the same way as above.   $|102\rangle$ is prepared by first making $|111\rangle$ and then performing the swap as described in Fig. 2.  Many additional eigenstates are close in energy but are irrelevant because they do not interact with the populated states.  A large avoided crossing between the relevant eigenstates  that is used to produce an adiabatic three-qubit interaction happens near $28~\mathrm{m}\Phi_0$.  Extra lines near $31~\mathrm{m}\Phi_0$ and $29~\mathrm{m}\Phi_0$ are due to third-order interactions predicted by the Hamiltonian, as is the larger first-order interaction at $25~\mathrm{m}\Phi_0$, but their effect on the protocol in Fig. 2 is negligible.
	}
	{\label{fig:chevrons}}
\end{figure}

Our three-qubit gate employs an interaction with the third excited state of one qubit.  Specifically, it relies on the unique capability among computational states ($\sigma_z$ eigenstates) of $\bra{111}$ (the notation $\bra{abc}$ refers to the excitation level of $Q_1$-$Q_3$, respectively) to interact with the non-computational state $\bra{003}$.  As the direct interaction of these states is first-order prohibited, we first transfer the quantum amplitude of $\bra{111}$ to the intermediate state $\bra{102}$, which itself couples strongly to $\bra{003}$.  Calculated energy levels and time-domain data showing interaction between $\bra{011}$ and $\bra{002}$ (which is identical to $\bra{111}$ and $\bra{102}$ except for a $6\ghz$ offset) as a function of the flux bias on $Q_2$ are shown in Fig. 1(a).  Once the amplitude of $\bra{111}$ is transferred to $\bra{102}$ with a sudden swap interaction, three-qubit phase is acquired by moving $Q_1$ up in frequency adiabatically, near the avoided crossing with $\bra{003}$.  Figure 1(b) shows the avoided crossing between these states as a function of the flux bias on $Q_1$.  This crossing shifts the frequency of $\bra{102}$ relative to the sum of $\bra{100}$ and $\bra{002}$ to yield our three-qubit phase.  The detailed procedure of the gate is shown in Fig. 2(a), taking a total of $63 \ns$.  Further details can be found in the Supplementary Information.

\begin{figure}
\centering
\includegraphics[scale=1]{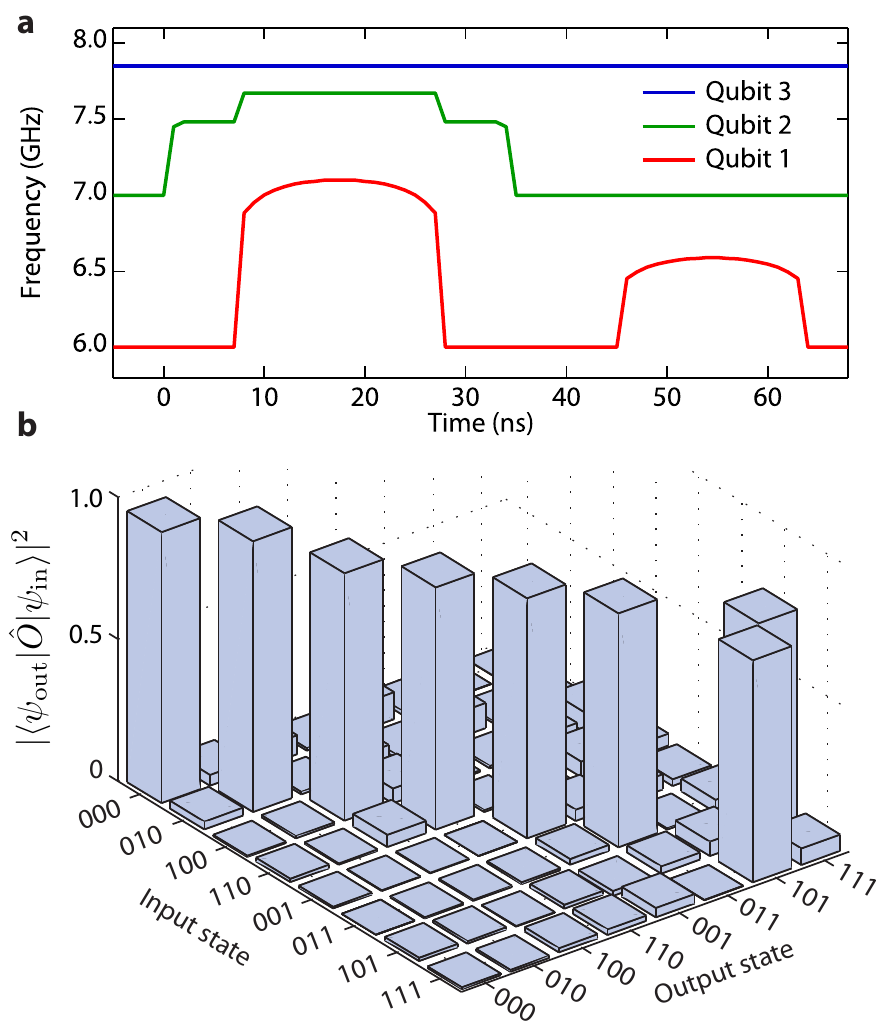}
\caption{
 	\textbf{Three-qubit gate pulse sequence and classical action.}
	(a) The frequency of the three qubits during the gate as a function of time.  First, $Q_2$ is moved suddenly into resonance with the avoided crossing shown in Fig. 1(a) to coherently transfer the population of $|111\rangle$ to $|102\rangle$ (and also $|011\rangle$ to $|002\rangle$) in $7\ns$.  Fine adjustments in the first point of the pulse compensates for finite pulse rise time and temporal precision.  $Q_2$ is then moved suddenly further up in frequency, to where its two-qubit phase with $Q_3$ is cancelled during the gate by accumulating a multiple of $2\pi$.  $Q_1$ is then moved up adiabatically to initiate the interaction between $|102\rangle$ and $|003\rangle$.  The duration and amplitude of this pulse is tuned to acquire a three-qubit phase of exactly $\pi$. The population in $|102\rangle$ is then transferred back to $|111\rangle$ by reversing the swap procedure.  Finally, the two-qubit phase between $Q_1$ and $Q_2$ is cancelled with an additional adiabatic interaction, which is sped up with a $\pi$ pulse on $Q_2$ at $37\ns$.  Here, this $\pi$ pulse is explicitly undone after the gate, but when it is used for error correction the following pulse is compiled together with other post-rotations.  The two-qubit phase between $Q_1$ and $Q_3$ is uncontrolled, making this a CC-$e^{i\phi}Z$ gate.
	(b) A CCNot gate is made by appending to the phase gate pre- and post-rotations on $Q_2$ as described in the Supplementary Information.  Its classical action is measured by preparing the eight computational basis states and performing state tomography on the result of applying the gate to them.  The projection of these measurements with the computational basis is taken to generate the truth table and is plotted.  The fidelity to the expected action, where only the states $|101\rangle$ and $|111\rangle$ are swapped, is $85 \pm 1\%$.
	}
{\label{fig:truthtable}}
\end{figure}

We first demonstrate the gate by measuring its classical action.  The controlled-controlled-phase (CCPhase) gate, which maps $\bra{111}$ to $-\bra{111}$, has no effect on pure computational states so we implement a CCNot gate by concatenating  pre- and post-rotations on $Q_2$, as described in the Supplementary Information.  Such a gate ideally swaps $\bra{101}$ and $\bra{111}$ and does nothing to the remaining states.  To verify this, we prepare the eight computational states, perform the gate, and measure its output with three-qubit state tomography \cite{DiCarlo2010} to generate the classical truth table.  The intended state is reached with $85 \pm 1\%$ fidelity on average.  This measurement is only sensitive to classical action, however, and a more thorough set of measurements is needed to fully characterize the gate.

\begin{figure*}
\centering
\includegraphics[scale=1]{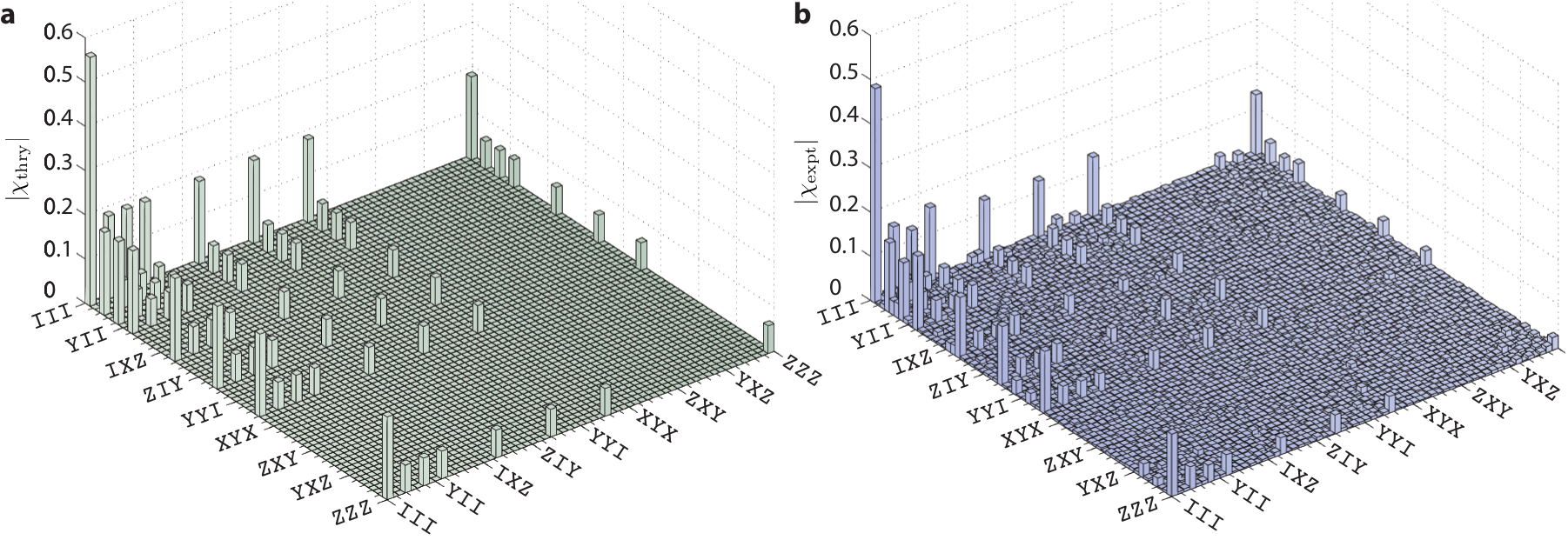}
\caption{
	\textbf{Quantum process tomography of the three-qubit phase gate.}
	Absolute values of the elements of the (a) ideal and (b) measured process matrices.  Data is collected by preparing 64 input states which span the three-qubit Hilbert space, applying the phase gate to them, and measuring the resulting density matrix with state tomography.  The process matrix $\chi$ of the operator $O$ is related to these data by $\rho_{\mathrm{out}} = O(\rho_{\mathrm{in}}) = \sum^{4^N}_{m,n=1} \chi_{mn} A_{m} \rho_{\mathrm{in}} A_{n}^{\dagger}$, where $\rho_{\mathrm{in}}$ is the density matrix of the input state, $\rho_{\mathrm{out}}$ is the measured output, $A_{i}$ is an operator basis spanning the three-qubit operator space, here chosen to be the tensor products of three Pauli matrices, and $N=3$ qubits \cite{Nielsen2000}.  The operator basis is ordered as in Ref. \citenum{DiCarlo2010} and is explicitly written in the Supplementary Information.  The ideal nonzero bars along the left edge are III, IIZ, IZI, ZII, IZZ, ZIZ, ZZI, and ZZZ.  The fidelity of the operation  $f=\mathrm{Tr}[\chi_{\mathrm{expt}} \chi_{\mathrm{thry}}] = 78 \pm 1$\%.  The ideal process matrix is calculated with the uncorrected phase between $Q_1$ and $Q_3$ set to its measured value of 57 degrees, which is irrelevant for our implementation of quantum error correction because the ancilla qubits would be reset to their ground state \cite{Reed2010} for a repeated cycle of correction.  The fidelity to the ``true'' CCPhase gate, where the $Q_1$-$Q_3$ phase is set to 0, is $69 \pm 1$\%.
	}
{\label{fig:qpt}}
\end{figure*}

To complete our verification, we perform full quantum process tomography (QPT) on the CCPhase gate.  In addition to detecting the action of the gate on quantum superpositions of computational states, QPT also detects non-unitary time evolution due to spurious coupling to the environment.  It is done by preparing 64 input states which span the computational Hilbert space and performing state tomography on the result of the gate's action on each state.  As shown in Fig. 3, the fidelity is found to be $78 \pm 1$\% to a process in which the spurious two-qubit phase between $Q_1$ and $Q_3$ is set to the measured value of 57 degrees (see the Supplementary Information for details on this phase and an explanation of why it is irrelevant here).  Due to this extraneous phase, the phase gate is most accurately described as a CC-$e^{i\phi}Z$ gate ($Z$ is a Pauli operator \cite{Nielsen2000}).  The infidelity is consistent with the expected energy relaxation of the three qubits during the $85 \ns$ measurement, with some remaining error owing to qubit transition frequency drift during the 90 minutes it takes to collect the full dataset.

\begin{figure}
\centering
\includegraphics[scale=1]{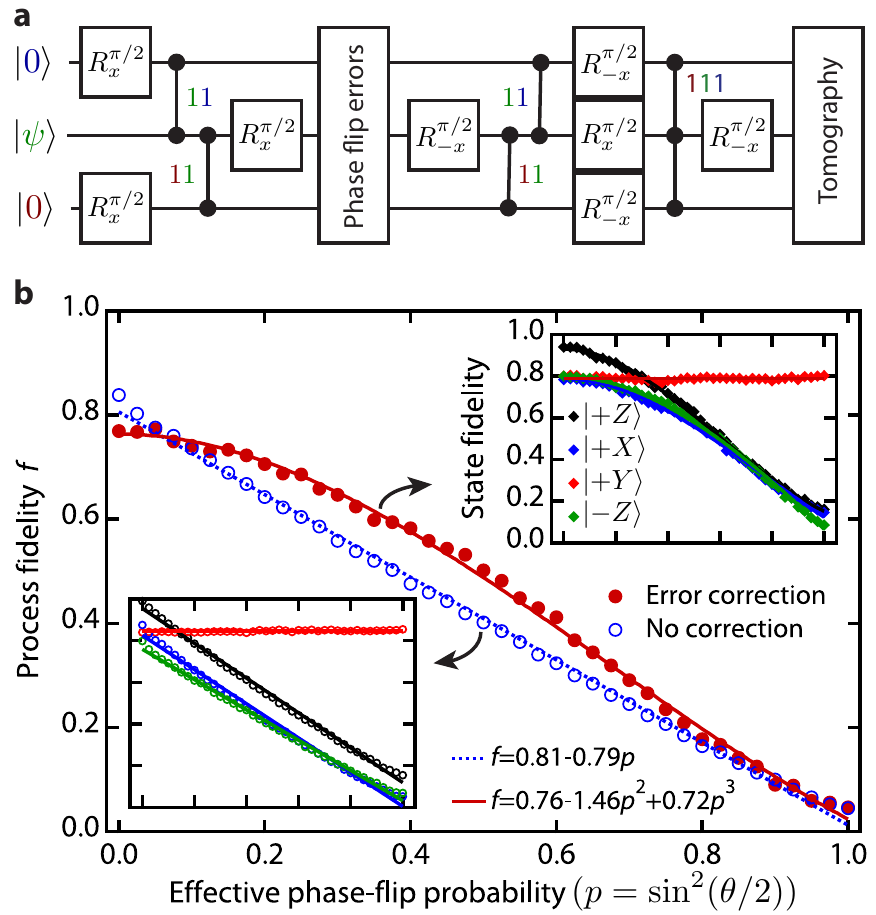}
\caption{
	\textbf{Three-qubit phase-flip error correction scheme and demonstration of first-order insensitivity to errors.}
	(a) The error correction protocol starts by entangling the two ancilla qubits with the primary qubit through the use of two CPhase gates (vertical lines terminating in solid circles).  The number adjacent to each indicates which state receives a phase shift.  A $\pi/2$ rotation on the primary qubit is then performed, making this a phase-flip error correction code.  If we wished to protect from bit flips, the two ancilla qubits would instead be rotated \cite{Nielsen2000}.  We perform errors on all three qubits simultaneously with $z$-gates of known rotation angle, which is equivalent to phase-flip errors with probability $p=\mathrm{sin}^2(\theta/2)$.  The encoding is then reversed, leaving the ancillas in a state indicating which single-qubit error occurred.  If an error has occurred on the primary qubit, the CCNot gate implemented with our CCPhase gate (represented by three solid circles linked by a vertical line) at the end of the code will reverse it.  We then perform three-qubit state tomography to measure the result.  
	(b) The fidelity of the protected qubit process matrix to the identity operation is plotted as a function of $p$.  As the code corrects only single-qubit errors, it will fail if more than one error occurs, which happens with probability $3p^2 - 2p^3$.  These coefficients are reduced for processes with finite fidelity.  The process fidelity is fit with $f = (0.76 \pm 0.005) - (1.46 \pm 0.03) p^2 + (0.72 \pm 0.03) p^3$.  If a linear term is allowed, its best-fit coefficient is $0.03 \pm 0.06$.  We compare this to the case of no error correction to simulate the improvement.
	(insets) The constituent state fidelities of the four basis states used to produce the process fidelity data for the case of (right) error correction and (left) no correction.  The state $\bra{\mathrm{+}Y}$ is immune to errors because its encoded state is an eigenvector of two-qubit phase flips.
	}
{\label{fig:qec}}
\end{figure}

With our CCPhase gate in hand, we now demonstrate three-qubit error correction.  Both the phase- and bit-flip codes begin by encoding the quantum state to be protected in a three-qubit entangled state \cite{Nielsen2000} by using conditional phase (CPhase) gates, as shown in Fig. 4(a).  The two codes differ only by single-qubit gates applied after entanglement in the encoding step.  For quantum states on the equator of the Bloch sphere, the resulting encoding is a maximally entangled three-qubit GHZ state \cite{Greenberger1989, DiCarlo2010, Neeley2010} which we independently measure to have a state fidelity of $89 \pm 1\%$.  Once the state is encoded, a single error of the chosen type on any of the qubits can be detected and corrected.  The error syndrome is decoded by reversing the encoding sequence, leaving the ancilla qubits ($Q_1$ and $Q_3$) in a state indicating which error occurred.  For a full flip, they will be in a computational state.  In particular, both ancillas will be excited if the primary qubit ($Q_2$) was flipped, and so the application of the CCNot gate will correct it.  As detailed in the Supplementary Information for the case of bit-flip errors, an arbitrary rotation on any single qubit about the protected axis can be also encoded, detected, and reversed.

In real physical systems, errors will occur at approximately the same rate on all constituent qubits.  The correction scheme will succeed, therefore, when the system projects to zero or one errors.  The probability of more than one error occurring is $3p^2 - 2 p^3$, where $p$ is the single-qubit error rate \cite{Nielsen2000}, and so the fidelity of error correction should be $1 - 3p^2 + 2 p^3$.  For a scheme with gate fidelity limited by decoherence, these coefficients will be smaller, but crucially, any linear dependence on $p$ will be strongly suppressed.  As shown in Fig. 4(b), we measure the process fidelity of the phase-flip error correction scheme as a function of $p$ by encoding four states which span the single-qubit Hilbert space and performing state tomography on the procedure's output.  We compare this to the case of no error correction in which identical single-qubit rotations are applied to $Q_2$ but the ancillas are not involved (and with appropriate delays to have the same total procedure duration).  Whereas without error correction we find a purely linear dependence on $p$, with the correction applied the data is extremely well modeled by only quadratic and cubic terms, demonstrating the desired first-order insensitivity to errors.

We have realized both bit- and phase-flip error correction in a superconducting circuit.  In doing so, we have tested both major conceptual components of the nine-qubit Shor code \cite{Shor1995b}, which can protect from arbitrary single-qubit errors by concatenating the bit- and phase-flip codes.  The implementation relies on our efficient three-qubit gate which employs non-computational states in the third excitation manifold of our system, demonstrating that the simple Hamiltonian of the system accurately predicts the dynamics even at these high excitation levels.  The gate takes approximately half the time of an equivalent construction with one- and two-qubit gates.  We expect it to work between any three nearest-neighbor qubits in frequency regardless of the number of qubits sharing the bus, as interactions involving other qubits will be first-order prohibited.  

We thank G. Kirchmair, M. Mirrahimi, I. Chuang, and M. Devoret for helpful discussions.  We acknowledge support from LPS/NSA under ARO Contract No.\ W911NF-09-1-0514 and from the NSF under Grants No.\ DMR-0653377 and No.\ DMR-1004406. Additional support provided by CNR-Istituto di Cibernetica, Pozzuoli, Italy (LF) and the Swiss NSF (SEN).

\section{Methods}

\subsection{Hamiltonian parameters}
The Tavis-Cummings Hamiltonian describing our system with four transmon qubits is
\begin{eqnarray}
\label{eq:Ham}
H &=& \hbar \wcav a^{\dag}a + \nonumber\\
  & & \hbar \sum_{q=1}^4 \Bigl( \sum_{j=0}^{N} \omega_{0j}^{(q)} \lvert j\rangle_q \langle j\rvert_q + (a+a^{\dag}) \!\!\sum_{j,k=0}^{N} g_{jk}^{(q)}\lvert j\rangle_q\langle k\rvert_q \Bigr). \nonumber
\end{eqnarray}
Here, $\hbar$ is Planck's reduced constant, $\wcav$ is the bare cavity frequency, $\omega^{(q)}_{0j}$ is the transition frequency for transmon $q$ from ground to excited state $j$, and $g_{jk}^{(q)}=g_q n_{jk}$, with  $g_q$ a bare qubit-cavity coupling and $n_{jk}$ a coupling matrix element.  $\omega^{(q)}_{0j}$ and $n_{jk}$ depend on transmon charging ($\ECq$) and Josephson ($\EJq$) energies \cite{Koch2007}.  Flux dependence comes from $\EJq=\EJqmax\lvert\cos(\pi\Phi_q/\Phi_0)\rvert$, with $\Phi_q$ the flux through the transmon SQUID loop and $\Phi_0$ is the flux quantum.  A linear flux-voltage relation $\Phi_q=\sum_{i=1}^4\alpha_{qi} V_i + \Phi_{q,0}$ describes crosstalk and offsets.  Spectroscopy and transmission data as a function of flux bias gives
$\wcav/2\pi=9.070~\GHz$,
$\EJqmax/h=\{33,35,26,57\}~\GHz$ (from $Q_1$ to $Q_4$), $g_q/2\pi \approx 220~\MHz$, and $\ECq/h\approx 330~\MHz$.  The measured qubit lifetimes for $Q_1$-$Q_3$ are $T_1 = (1.3, 0.9, 0.7)~\us$ and coherence times $T_2^* = (0.5, 0.6, 1.3)~\us$ respectively.

\subsection{Qubit rotations and gate calibration}
Arbitrary qubit rotations around the $x$- and $y$-axis of the Bloch sphere are performed with pulse-shaped resonant microwave tones.  Rotations around the $z$-axis are done by rotating the reference phase of subsequent $x$ and $y$ pulses.  One-qubit dynamical phases resulting from flux excursions are measured with modified Ramsey experiments comparing the phase difference between an unmodified prepared state and that same state after a flux pulse and are cancelled with $z$ rotations.  Two- and three-qubit phases are measured with a similar Ramsey experiment comparing the phase difference acquired when a control qubit is in its ground and excited state.  For example, the two-qubit phase between $Q_2$ and $Q_3$ is measured by preparing $Q_3$ along the $y$-axis and $Q_2$ either in its ground or excited state and then performing the flux pulse in both cases.  The single-qubit phase of $Q_3$ is the same for both states, and so the two-qubit phase is directly measurable as their phase difference.  All phases are initially tuned to within one degree, limited by the resolution of control equipment and drifts of system parameters such as the qubit transition frequencies.

\bibliography{toffoli}

\end{document}


\title{Supplementary Information for ``Realization of Three-Qubit Quantum Error Correction with Superconducting Circuits''}

\author{M. D. Reed}
\affiliation{Departments of Physics and Applied Physics, Yale University, New Haven, Connecticut 06520, USA}

\author{L. DiCarlo}
\affiliation{Kavli Institute of Nanoscience, Delft University of Technology, Delft, The Netherlands}

\author{S. E. Nigg}
\affiliation{Departments of Physics and Applied Physics, Yale University, New Haven, Connecticut 06520, USA}

\author{L. Sun}
\affiliation{Departments of Physics and Applied Physics, Yale University, New Haven, Connecticut 06520, USA}

\author{L. Frunzio}
\affiliation{Departments of Physics and Applied Physics, Yale University, New Haven, Connecticut 06520, USA}

\author{S. M. Girvin}
\affiliation{Departments of Physics and Applied Physics, Yale University, New Haven, Connecticut 06520, USA}

\author{R. J. Schoelkopf}
\affiliation{Departments of Physics and Applied Physics, Yale University, New Haven, Connecticut 06520, USA}

\date{\today}

\ifpdf
\DeclareGraphicsExtensions{.pdf, .jpg, .tif}
\else
\DeclareGraphicsExtensions{.eps, .jpg}
\fi

\maketitle

\section{Details of three-qubit phase gate}

To understand the physical mechanism behind our three-qubit gate, it is useful to first review how two-qubit CPhase gates are commonly implemented in the cQED architecture.  The avoided crossing between the first excited state of two transmon qubits ($\bra{11}$) and the second excited state of one ($\bra{02}$) can be employed both adiabatically \cite{DiCarlo2009} or suddenly \cite{DiCarlo2010, Strauch2003} to produce a CPhase gate.  If the system is adiabatically tuned to the vicinity of the avoided crossing, the system will remain fully in the eigenstate that maps to the computational subspace away from the crossing. However, the interaction causes the energy level of the $\bra{11}$ state to differ from the sum of the $\bra{01}$ and $\bra{10}$ states, advancing the quantum phase of $|11\rangle$ relative to the quantum phases of $\bra{10}$ and $\bra{01}$, and entangling the qubits \cite{DiCarlo2009}.  Alternatively, if the avoided crossing is small enough, the gate can be executed suddenly.  At the avoided crossing, the eigenbasis consists of the symmetric ($\bra{+}$) and antisymmetric ($\bra{-}$) superpositions of $\bra{11}$ and $\bra{02}$.  Starting from $\bra{11}$ and moving suddenly, the wavefunction will no longer be in an energy eigenstate, but rather an equal superposition of $\bra{+}$ and $\bra{-}$ whose relative phase will advance with the interaction strength, oscillating between the undressed states $\bra{11}$ and $\bra{02}$.  Waiting one full period will return the state to the computational basis, but with a phase difference of $\pi$, at which point it can be moved suddenly away from the avoided crossing to return to the undressed eigenbasis.

\begin{figure*}
\centering
\includegraphics[scale=1]{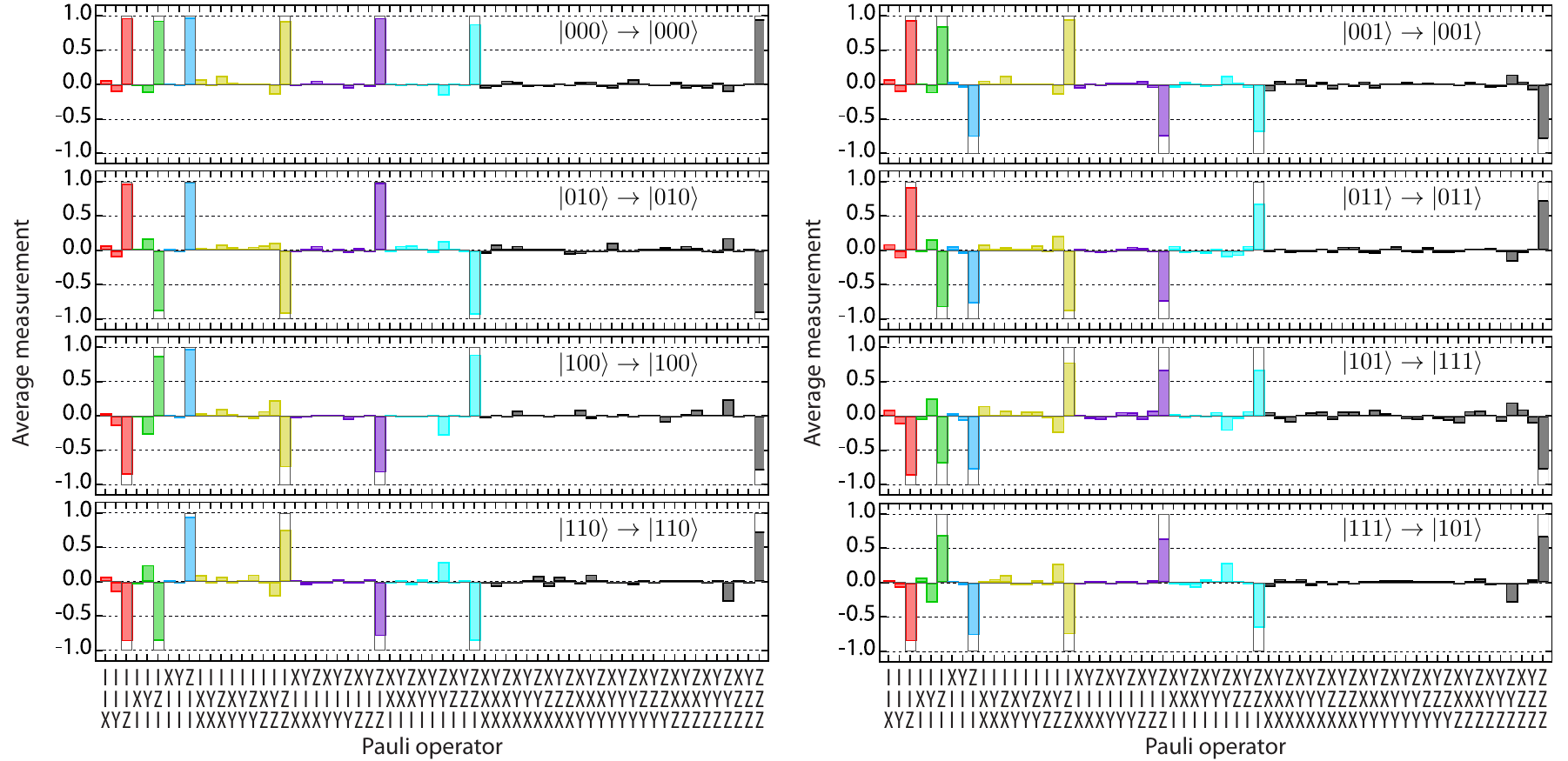}
\caption{
 	\textbf{Reconstructed density matrices of the result of applying the CCNot gate to computational states.}  Each computational state is prepared and the CCNot gate described below is acted on it.  Ideally, all states would have nothing happen to them except for $\bra{101}$ and $\bra{111}$, which swap.  The computational state prepared and what it should map to is indicated on each tomogram, which is visualized as the Pauli set as found in Ref. \citenum{DiCarlo2010}.  The Pauli set of the ideal state is superposed (open bars) and is reached with fidelity 95.4, 95.2, 87.1, 85.2, 84.4, 82.2, 78.1, and 75.7\% for each of the states, respectively.  Note that the lack of spurious two- and three-qubit correlations indicates that there is no significant loss of population from the computational subspace.
	}
{\label{fig:truthtabletomo}}
\end{figure*}

Any two-qubit phase gate can be described in terms of one- and two-qubit phases, which is a helpful abstraction to understand the different sources of phase delay.  In this language, a two-qubit phase gate maps $|00\rangle \rightarrow |00\rangle$, $|10\rangle \rightarrow e^{i \phi_{10}} |10\rangle$, $|01\rangle \rightarrow e^{i \phi_{01}} |01\rangle$, and $|11\rangle \rightarrow e^{i (\phi_{10}+\phi_{01}+\phi_{11})} |11\rangle$.  Here $\phi_{10}$ and $\phi_{01}$ are single-qubit phases given by the time-integrated detuning of a qubit transition frequency from its nominal bias point, and $\phi_{11}$ is a two-qubit phase which can be generated as described above.  Note that the state $\bra{11}$ suffers phases from all of these sources, and so measuring the two-qubit interaction can only be done with separate measurements of $\phi_{01}$ and $\phi_{10}$.  Combining an interaction generating $\phi_{11}=\pi$ with appropriate one-qubit phases can yield a CPhase gate conditioned on any target computational state \cite{DiCarlo2009}.

The three-qubit phase of our CCPhase gate arises from an interaction between $\bra{111}$ and $\bra{003}$, in analogy to the two-qubit case.  In the same way that the CPhase gate requires two excited qubits to access $\bra{02}$, the CCPhase gate requires three excited qubits to access $\bra{003}$.  The direct interaction between these states is first-order prohibited because the states differ by a change of two excitations, so we achieve an effective interaction by using a state which couples strongly to both: $\bra{102}$.  As described in the main text, the gate is initiated with a full coherent transfer of the population of $\bra{111}$ to $\bra{102}$ (and $\bra{011}$ to $\bra{002}$) by suddenly approaching the crossing and oscillating between the two dressed eigenstates for exactly half the splitting period before moving the qubit suddenly further up in frequency.  Small errors due to imperfect timing and finite pulse rise time are corrected by shaping the qubit's trajectory to minimize residual $\bra{002}$ population after swapping into the state and back.  Thanks to the strong coupling ($67 \mhz$) of these states, the transfer takes only $7 \ns$.  Once the  amplitude of the target state is transferred to $\bra{102}$, the three-qubit phase is acquired by moving $Q_1$ up in frequency, near the avoided crossing with $\bra{003}$.  The interaction strength between these states is large ($121\mhz$) so we choose to acquire this three-qubit phase adiabatically.  As in the case of two qubits described above, $\bra{102}$ experiences a frequency shift relative to its constituents ($\bra{100}$ plus $\bra{002}$) due to the avoided crossing, yielding the three-qubit phase.

\begin{figure*}
\centering
\includegraphics[scale=1]{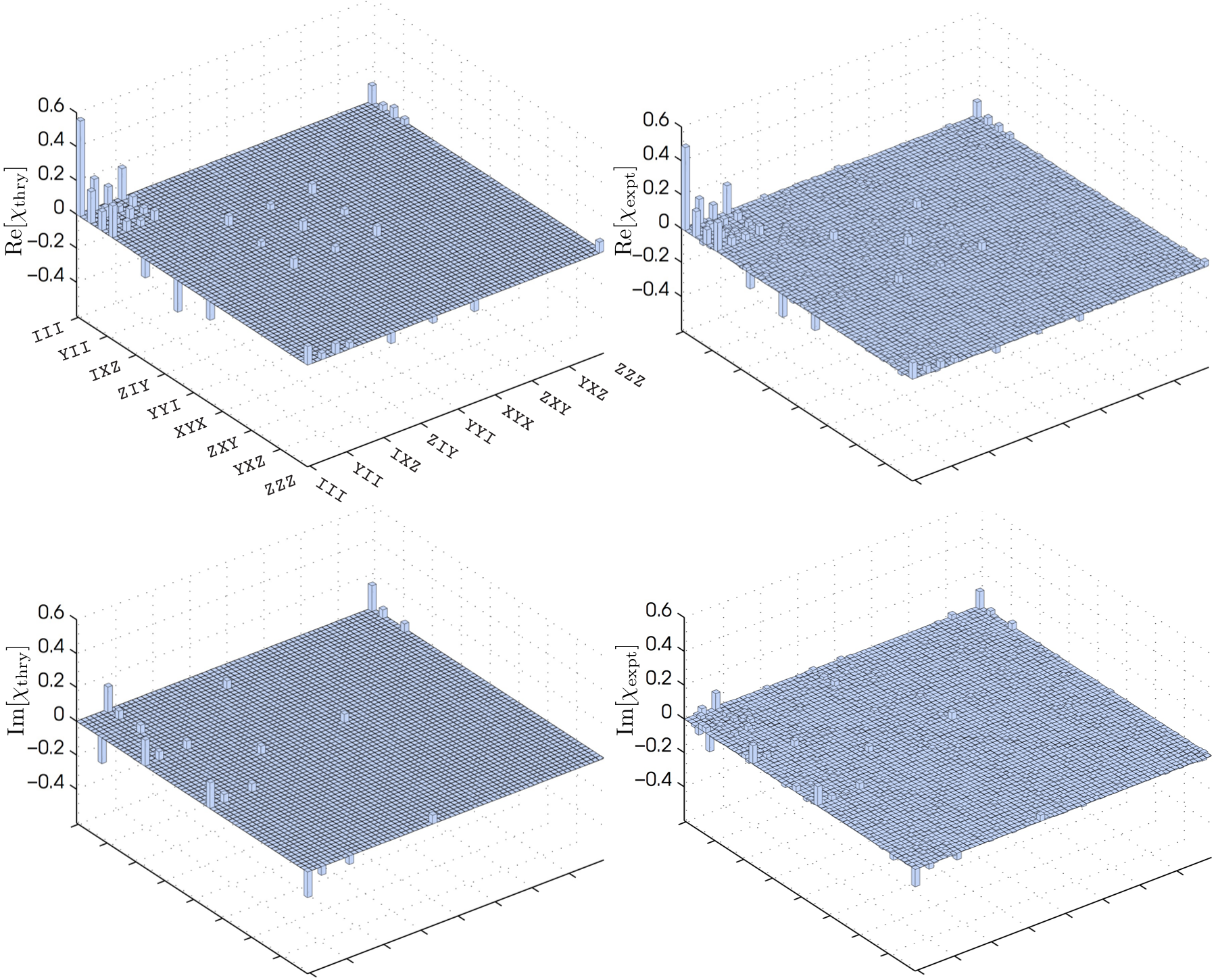}
\caption{
 	\textbf{Real and imaginary parts of ideal and measured process tomography $\boldsymbol{\chi}$ matrix.}  These data were collected as described in Fig. 3 of the main text.  There only the absolute value is shown, but here the full real and imaginary parts are reproduced.  The order of operators here and in Fig 3. of the main text is as follows: III, IIX, IIY, IIZ, IXI, IYI, IZI, XII, YII, ZII, IXX, IYX, IZX, IXY, IYY, IZY, IXZ, IYZ, IZZ, XIX, YIX, ZIX, XIY, YIY, ZIY, XIZ, YIZ, ZIZ, XXI, YXI, ZXI, XYI, YYI, ZYI, XZI, YZI, ZZI, XXX, YXX, ZXX, XYX, YYX, ZYX, XZX, YZX, ZZX, XXY, YXY, ZXY, XYY, YYY, ZYY, XZY, YZY, ZZY, XXZ, YXZ, ZXZ, XYZ, YYZ, ZYZ, XZZ, YZZ, and ZZZ.  We do not make use of the maximum-likelihood estimator commonly used to require the physicality of $\chi$ matrix so that the reported elements of $\chi$ and the fidelity are linearly related to the raw measurements \cite{Chow2010b}.  The uncertainty of the fidelities reported in the main text is given by the standard deviation of six repeated measurements of the full process matrix.
	}
{\label{fig:chimatrix}}
\end{figure*}

Again extending the two-qubit case, a three-qubit phase gate can be parametrized with seven unique phases.  Three are one-qubit phases ($\phi_{001}$, $\phi_{010}$, and $\phi_{100}$), three are two-qubit phases ($\phi_{011}$, $\phi_{101}$, and $\phi_{110}$), and one is a three-qubit phase ($\phi_{111}$).  Ideally, our gate procedure would provide full control over them all, but we make a simplifying assumption based on the intended application of the gate.  During error correction, errors are transferred from the protected qubit to the ancillas, which are then reset either through measurement-conditioned pulses or by coupling them to a dissipative bath \cite{Reed2010}.  As their final state does not matter, we are free to choose any one two-qubit phase to remain uncorrected: here, the non-nearest neighbor interaction given by $\phi_{101}$, which is most challenging to control.  This implies that $Q_2$ will be the target of our error correction scheme, with $Q_1$ and $Q_3$ acting as the ancillas.  The remaining two-qubit phases $\phi_{011}$ and $\phi_{110}$, however, must be set to zero.  The former can be easily corrected via fine-tuning of the frequency of $Q_2$ while in the $\bra{x02}$ state ($x=0,1$), where it acquires that phase very rapidly and so can be made to be an integer multiple of $2\pi$.  The latter phase, however, must be explicitly corrected with an additional adiabatic phase gate.  Because this angle is small, it is advantageous to $\pi$-pulse $Q_2$ prior to the adiabatic interaction, reversing the direction of phase evolution and reducing the overall correction time.  The action of the gate is to set $\phi_{100}$, $\phi_{010}$, $\phi_{001}$, $\phi_{110}$, and $\phi_{011}$ to zero, $\phi_{111}$ to $\pi$, with $\phi_{101}$ measured to be 57 degrees.  

A CCNot gate is constructed by appending a $\pi/2$ and a $-\pi/2$ pulse on $Q_2$ before and after the phase gate described in Fig. 2(a) of the main text.  For the two input states where both ancillas ($Q_1$ and $Q_3$) are excited, there is an effective $\pi$ phase shift applied to the second pulse, and so the two pulses add together to a full rotation.  Other states are not shifted so the pulses cancel.  The full state tomograms of the output of CCNot with the eight computational states as input are shown in Fig. S1.  These tomograms are used to derive the classical truth table shown in Fig. 2(b) of the main text.

All relevant phases are controlled to one degree or better, with their accuracy set by the voltage resolution of our arbitrary waveform generator.  This implies a ``quantum phase fidelity'' as defined in a recent work \cite{Mariantoni2011} in excess of 99\% to the six relevant phases.  This metric is not very sensitive to even phase errors, however, and so true quantum process fidelity should be used whenever possible.  For example, the phase fidelity of an ideal (e.g. decoherence-free) CCPhase gate with a maximal single qubit phase error of $\phi_{100} = \pi$ to a CCPhase in which $\phi_{100}=0$ is $85.7\%$ despite the fact that the quantum process fidelity between those gates reveals them to be nearly orthogonal, with a fidelity of $6.3\%$.  The real and imaginary parts of the measured process matrix are shown in Fig. S2.

\section{Bit-flip error correction}

In order to illuminate the error projection process, we demonstrate bit-flip error correction with errors on only one qubit at a time.  As in the phase-flip case described in the main text, bit-flip correction begins by encoding a quantum state in a three-qubit entangled state \cite{Nielsen2000} using two sudden CPhase gates paired with appropriate single-qubit rotations \cite{DiCarlo2010}.  The difference between the phase- and bit-flip codes lies only in the rotations performed after the entanglement.  Instead of rotating the primary qubit as in phase-flip correction, the ancillas are $\pi/2$ pulsed.  For simplicity, here we only measure state tomography (rather than process tomography) of the state most sensitive to bit-flip errors of the chosen type: the positive eigenstate of $\sigma_x$.  The state is now in a protected subspace which can recover from any single spurious $y$ rotation on any of the three qubits.

\begin{figure}
\centering
\includegraphics[scale=1]{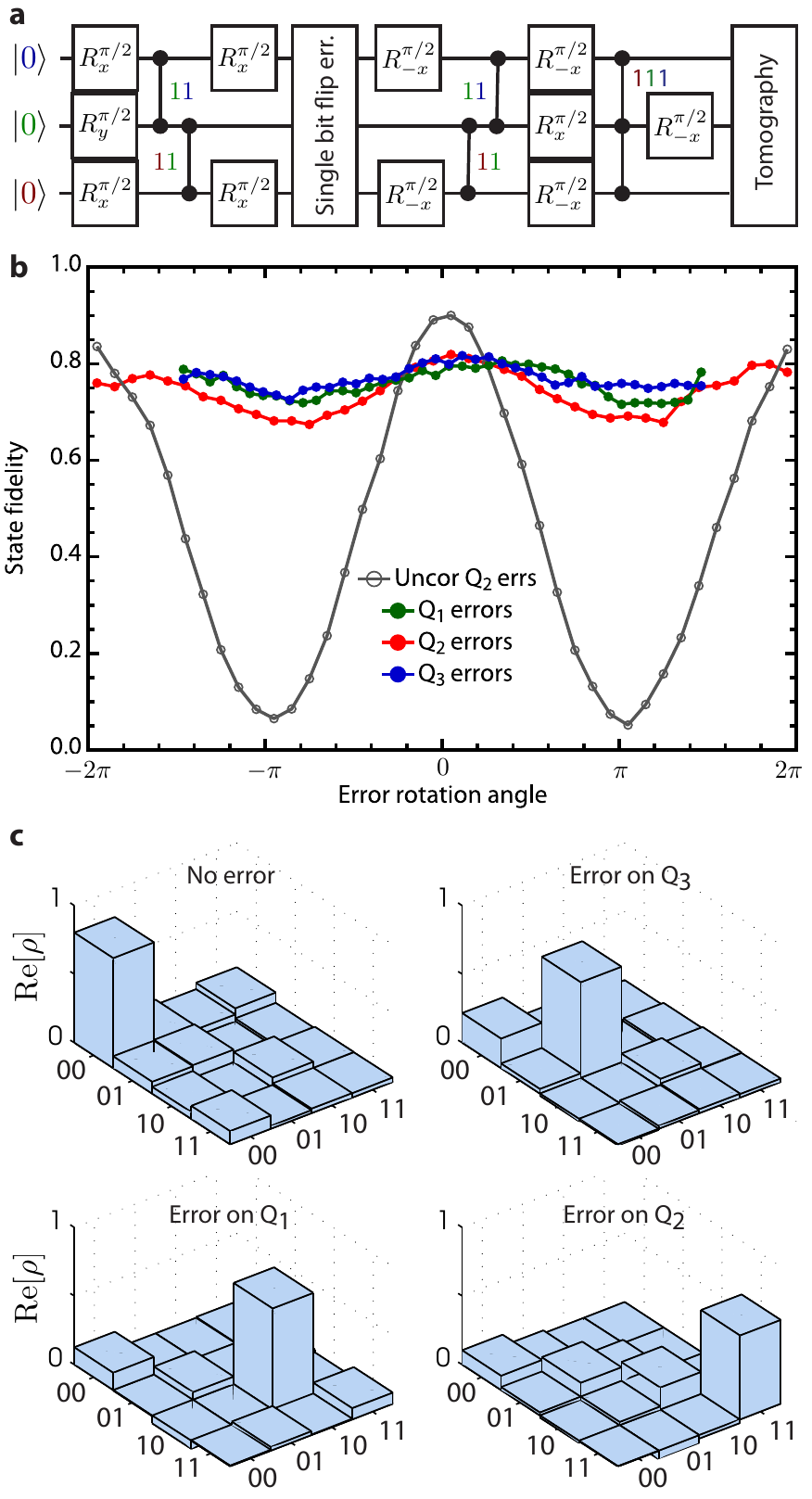}
\caption{
 	\textbf{Bit-flip error correction.}  (a) Bit-flip error correction gate sequence.  This differs from Fig. 4(a) of the main text by only single-qubit rotations.  Note also that for simplicity here we do only state tomography on the output of the process on the maximally affected state, so the state preparation is as indicated for $Q_2$.  (b) State fidelity to the created state after performing a single error on only one of the qubits, with and without error correction.  Ideally, the curves would be flat lines at unit fidelity.  Finite excited-state lifetimes cause oscillations and displacement down as the errors change the excitation level of the system.  (c) Two-qubit density matrices of the ancillas after each of the four possible errors has occurred.  The fidelity of each of these states to the ideal error syndromes are $(81.3\%, 69.7\%, 73.1\%, 61.2\%)$.
	}
{\label{fig:bitflip}}
\end{figure}

We perform intentional rotations on one of the qubits with a varying rotation angle $\theta$.  In the framework of error correction \cite{Nielsen2000}, errors are decomposed as probabilistic full bit flips rather than continuous rotations, and so our partial rotations can be instead seen as varying the probability of a full bit flip.  In the case of the ancillas, this $y$-rotation would normally be placed between a positive and negative $\pi/2~x$-rotation associated with turning a CPhase into a CNot, and so we compile these three single-qubit gates into one $z$-gate.  After the error has occurred, we disentangle the three-qubit state, effectively encoding an error code in the ancillas.  This code will leave both ancillas excited if and only if a bit flip has occurred on the primary qubit, which is then reversed by our CCNot gate.  At this point, the entropy of the error is stored in the ancillas, which should be reset via coupling to a cold bath \cite{Reed2010} if we were to loop the code.  In Fig. S3(b), we plot the fidelity of the protected qubit to its prepared state after the error correction procedure has been applied to errors on all three qubits and also if no error correction is done.  Ideally, these curves would be flat and with unit fidelity, but because of qubit decay and the varying excitation level of the qubits depending on the error performed, the curves show a small oscillation centered at 75.7\% fidelity. 

We also measure the density matrices of the ancilla qubits after the four possible full bit-flip errors (no error and bit flips on each of the three qubits).  The code should ideally encode each of those four possible errors as one of the four computational states of the two-qubit ancilla subspace.  As shown in Fig. S3(c), the measured density matrices of the ancillas do indeed encode the error syndrome as expected, albeit with finite fidelity.  The measured state fidelity to the ideal syndrome is $f_{\mathrm{syndrome}} = (81.3\%, 69.7\%, 73.1\%, 61.2\%)$ to the states $(\bra{00}, \bra{10}, \bra{01}, \bra{11})$ encoding no error and errors on $Q_3$, $Q_1$, and $Q_2$ respectively.  In the case of a finite rotation, the ancillas will instead be in a superposition of error and no error, and so the gate will coherently correct the primary qubit, acting only on the subspace where the implicated error occurred.  This action is free to be done by measurement and conditional feed-forward pulses (that is, by projecting the qubits with measurements), however, and indeed simplifies the requirements for fault-tolerance if it is done that way \cite{Nielsen2000}.

\bibliography{toffoli}